\documentclass[aps,prd,floats,floatfix,nofootinbib]{revtex4-1}
\usepackage{graphicx,amsmath}
\usepackage{amssymb}
\usepackage{amsfonts}
\usepackage{hyperref}
\usepackage[bottom]{footmisc}

\begin{document}

\title{A Flavorful Top-Coloron Model}

\author{R.\ Sekhar Chivukula}
\email {sekhar@msu.edu}
\affiliation {Department of Physics and Astronomy,
Michigan State University,
East Lansing, MI 48824, USA}

\author{Elizabeth H.\ Simmons}
\email {esimmons@pa.msu.edu}
\affiliation {Department of Physics and Astronomy,
Michigan State University,
East Lansing, MI 48824, USA}

\author{Natascia Vignaroli}
\email {vignaroli@pa.msu.edu}
\affiliation {Department of Physics and Astronomy,
Michigan State University,
East Lansing, MI 48824, USA}

\date{\today}

\begin{abstract}
In this paper we introduce a simple renormalizable model of an extended color gauge sector
in which the third-generation quarks couple differently than the lighter quarks.  In
addition to a set of heavy color-octet vector bosons (colorons), the model also contains a set of 
heavy weak vector quarks. Mixing between the third-generation of quarks and
the first two is naturally small, and occurs only through the (suppressed) mixing of all 
three generations with the heavy vector
quarks. We discuss the constraints on this
model arising from limits on flavor-changing neutral currents and from collider searches
for the colorons and vector quarks, and discuss the prospects for discovery at
the LHC.
\end{abstract}

\maketitle

\section{Introduction}

The LHC Era has started off with a bang: not only have the experiments decisively rediscovered all of the familiar particles of the Standard Model, confirming that the operations of the acclerator and detectors are well understood, but both ATLAS \cite{ATLAS:2012gk} and CMS \cite{:2012gu} have also found a new scalar particle that closely resembles the long-awaited Higgs Boson. Since the Standard Model with one Higgs doublet is not natural up to arbitrarily high energies, and since it leaves many questions (including the origin of flavor) unanswered, we anticipate that some physics beyond the Standard Model remains to be discovered. 

An intriguing possibility is that an extended color gauge sector may exist.  In particular, new colored states beyond the familiar quarks and gluons could be awaiting discovery at the LHC.  These could reflect a variety of kinds of theories beyond the standard model.  One class of theories are those in which the strong interactions are extended from the standard $SU(3)_{QCD}$ to a larger $SU(3)_1 \times SU(3)_2$ group and in which spontaneous symmetry breaking reduces the larger group to its diagonal subgroup which is identified with $SU(3)_{QCD}$.  These models include topcolor \cite{Hill:1991at}, the flavor-universal topcolor \cite{Chivukula:1996yr}, classic chiral color \cite{Frampton:1987dn}, chiral color with unequal gauge couplings \cite{Martynov:2009en} and a newer flavor non-universal chiral color model \cite{Frampton:2009rk}. Each of these models includes new heavy colored gauge bosons (colorons, topgluons, or axigluons) transforming as a color octet.  Other theories with new color octet states include theories of new extra spacetime dimensions that incorporate Kaluza-Klein partners for the gluons, as in Refs.  \cite{Dicus:2000hm}\cite{Davoudiasl:2000wi}\cite{Lillie:2007yh} or technicolor models with colored technifermions that bind into color-octet techni-rho mesons \cite{Farhi:1980xs}.  An entire catalog of possible new colored states including color sextet fermions, colored scalars, and low-scale string resonances \cite{Antoniadis:1990ew} has also been reviewed in \cite{Han:2010rf}.

If an extended color gauge sector does exist, then there are indications that it could be likely to couple more strongly to the third generation than the light quarks.  For instance, if the new scalar state with a mass of 125 GeV is actually a composite, rather than a fundamental, scalar, it could potentially be a bound state of top quarks\cite{Miransky:1988xi}\cite{Miransky:1989ds}\cite{Nambu:1989jt} \cite{Marciano:1989xd}\cite{Bardeen:1989ds}, as realized in topcolor \cite{Hill:1991at}, topcolor-assisted technicolor \cite{Hill:1994hp} and top seesaw \cite{Dobrescu:1997nm}\cite{Chivukula:1998wd}\cite{He:2001fz} models, and as analyzed phenomenologically in \cite{Chivukula:2012cp}\cite{Barbieri:2012tu}. There is also the puzzling question of how to explain the forward-backward asymmetry observed by the Tevatron experiments \cite{Aaltonen:2011kc}\cite{Abazov:2011rq} in the production of top-quark pairs.  Models involving flavor non-universal axigluons \cite{Frampton:2009rk} have been cited as a possible explanation, and there has been discussion in the literature \cite{Chivukula:2010fk}\cite{Bai:2011ed} of the degree to which the properties of those axigluons would be constrained by data on flavor-changing neutral currents.

In this paper, we introduce a model in which the strong interactions are extended to an $SU(3)_1 \times SU(3)_2$ structure in a way that causes the new heavy coloron states to couple differently to the third generation quarks than to the lighter generations.  What is novel about this model is that it also naturally addresses the experimental observation that the third family of quarks has only a small mixing with the lighter families.  This is implemented through the presence of heavy weak-vector quarks that transform in the same way under the extended color group as the third-generation quarks. Mixing between the ordinary quark generations occurs only because all three generations mix with the vector quarks; the different gauge charges of the third and lighter generations of quarks thus lead to naturally small mixing between those generations.  Effectively, this model is a nice realization of next-to-minimal flavor violation 
\cite{Agashe:2005hk,Barbieri:2011ci, D'Agnolo:2012ie, Barbieri:2012uh, Barbieri:2012bh, Buras:2012sd}.

In the next section, we will introduce the gauge and fermion sectors of the model, discuss the flavor symmetries, and demonstrate that the model naturally reproduces the CKM structure of quark masses and mixings.  In section 3, we study how data on flavor-changing neutral currents (FCNC) constrains the model parameters and find that clear regions remain allowed.  Section 4 shows how the LHC experiments' searches for new resonances decaying to dijets bound the properties of the colorons in our model; section 5, likewise, shows how LHC searches for new heavy colored fermions restrict the characteristics of the heavy vector quarks in our model.  Finally, in section 6, we summarize our conclusions and findings.

\section{The Model}

We will now introduce the model in more detail.  First, we discuss the gauge boson, scalar, and fermion content.   Then we detail the Dirac and Yukawa interaction terms and the flavor symmetries that yield the diverse masses of the nearly-standard fermion states.  Finally, we obtain explicit expressions for the fermion mass eigenstates and demonstrate that the observed pattern of masses and CKM mixings is obtained for natural values of the model parameters. 

\subsection{Gauge Structure}

We investigate a simple, renormalizable, model with the gauge structure 
$SU(3)_1 \times SU(3)_2 \times SU(2)_W \times U(1)_Y$. 
We name the $SU(3)_1 \times SU(3)_2$ gauge bosons $A^a_{1\mu}$ and $A^a_{2\mu}$, respectively, and call the 
corresponding gauge couplings $g_1$,  $g_2$. The two $SU(3)$ gauge couplings are related to the QCD coupling $g_S$ through
\begin{equation}
g_S=g_1\sin\omega=g_2\cos\omega ,
\end{equation}
where $\omega$ is a new gauge mixing angle.

Gauge symmetry breaking occurs in two steps:
\begin{itemize}
\item $SU(3)_1 \times SU(3)_2 \to SU(3)_C$ due to the (diagonal) expectation value
$\langle \Phi \rangle \propto u \cdot {\cal I}$, where the scalar, $\Phi$, transforms as a $(3,\bar{3})$ under
$SU(3)_1 \times SU(3)_2$ and ${\cal I}$ is the identity matrix,

\item $SU(2)_W \times U(1)_Y \to U(1)_{em}$ in the usual way due to a Higgs doublet
$\phi$ transforming as a $2_{1/2}$ of the electroweak group, and with the usual
vacuum expectation value given by $v \approx 246$ GeV.
\end{itemize}
We will assume that the color-group symmetry breaking occurs at a scale large compared
to the weak scale, $u \gg v$.\footnote{The vacuum expectation values for $\phi$ and $\Phi$
occur for a choice of parameters in the most general, renormalizable, potential for these fields, and
the vacuum is unique up to an arbitrary global gauge transformation. We will assume that the additional
physical singlet and color-octet fields in $\Phi$ are heavy, and neglect them in what follows \cite{Bai:2010dj}.}

The mass-squared matrix for the colored gauge bosons is given by
\begin{equation}
-\dfrac{1}{2} u^2\begin{pmatrix} g_1^2& -g_1 g_2\\ -g_1 g_2& g_2^2 \end{pmatrix}\, .
\end{equation}
Diagonalizing this matrix reveals mass eigenstates $G^a$ and $C^a$
\begin{eqnarray}
G^a_\mu &=& \sin\omega A^a_{1\mu} + \cos\omega A^a_{2\mu} \\
C^a _\mu &=& \cos\omega A^a_{1\mu} - \sin\omega A^a_{2\mu} \label{eq:coloron}
\end{eqnarray}
with masses
\begin{equation}
M_G = 0 \qquad M_C = u \sqrt{g_1^2 + g_2^2} = \dfrac{g_S\, u}{\sin\omega \cos\omega}
\end{equation}

If we name the color current associated with $SU(3)_i$ by the symbol $J_i^{a\mu}$, then the gluon and coloron, respectively couple to the following currents:
\begin{eqnarray}
g_S J_G^{a\mu} &=& g_S (J_1^{a\mu} + J_2^{a\mu})\\
g_S J_C^{a\mu} &=& g_S (\cot\omega J_1^{a\mu} - \tan\omega J_2^{a\mu})
\label{eq:coloroncurrent}
\end{eqnarray}
From this, we calculate the decay width of the coloron into massless color-triplet fermions to be 
\begin{equation}
\Gamma_C = \dfrac{g_S^2 M_c}{24 \pi} \left(n_1 \cot^2\omega + n_2 \tan^2\omega \right)
\end{equation}
where $n_1$ and $n_2$ correspond to the number of Dirac fermion states charged under $SU(3)_1$ and $SU(3)_2$ 
respectively. Finally, we note that at energy scales well below the coloron mass, coloron exchange  may be approximated by the current-current interaction:
\begin{equation}
-\dfrac{g^2_S}{2 M_c^2} J_C^{a\mu} J_{C\mu}^a\,.
\end{equation}

\subsection{Matter Fields}

\begin{table}
\centering
\begin{tabular}{|c|c|c|c|c|}
\hline
Particle & $SU(3)_1$ & $SU(3)_2$ & $SU(2)$ & $U(1)$ \\
\hline\hline
$\vec{\mathcal Q}_L= \begin{pmatrix}
 q_L \\
 Q_L
 \end{pmatrix}$ & 3 & 1 & 2 & +1/6\\
\hline
$t_R$ & 3 & 1 & 1 & +2/3\\
\hline
$b_R$ & 3 & 1 & 1 & -1/3\\
\hline
$Q_R$ & 3 & 1 & 2 & +1/6\\
\hline\hline
$\vec{\psi}_L = \begin{pmatrix}
 \psi^1_L \\
 \psi^2_L
 \end{pmatrix}
$ & 1 & 3 & 2 & +1/6\\
\hline
$\vec{u}_R= \begin{pmatrix}
 u^1_R \\
 u^2_R
 \end{pmatrix}$ & 1 & 3 & 1 & +2/3\\
\hline
$\vec{d}_R= \begin{pmatrix}
 d^1_R \\
 d^2_R
 \end{pmatrix}$ & 1 & 3 & 1 & -1/3\\
\hline\hline
$\phi$ & 1 & 1 & 2 & +1/2\\
\hline
$\Phi$ & 3 & $\bar{3}$ & 1 & 0 \\
\hline\hline
\end{tabular}
\caption{$SU(3)_1 \times SU(3)_2 \times SU(2) \times U(1)$ gauge charges of
the particles in this model. The $\phi$ and $\Phi$, respectively, denote the scalars responsible
for breaking the electroweak and (extended) strong sectors, while all other listed 
particles are fermions. The vectors ($\vec{\psi}_L$, $\vec{u}_R$,
$\vec{d}_R$, and $\vec{\mathcal Q}_L$) denote different fermion flavors with the
same gauge charges, where the superscripts [1,2] refer to the two light fermion generations. \label{table:i}}
\end{table}

The matter fields of this model are summarized\footnote{The lepton fields are assigned
to $SU(2) \times U(1)$ just as in the standard model. We normalize hypercharge such
that $Q=T_3 + Y$.}  in Table \ref{table:i}. Those coupled to $SU(3)_1$ include one chiral 
quark generation ($q_L$, $t_R$, and $b_R$), which will be associated primarily with
the third generation quarks, and one vectorial quark generation ($Q_{L,R}$). The two remaining (chiral)
quark generations ($\vec{\psi}_L$, $\vec{u}_R$, and $\vec{d}_R$) are coupled to $SU(3)_2$ and
will be associated primarily with the two light quark generations. 
Noting that $Q_L$ and $q_L$ transform in the same way under the gauge symmetries, we  define $\vec{Q}_L \equiv (q_L, Q_L)$ and observe that the flavor symmetries (ignoring 
gauge anomalies) of the quark kinetic energy terms in this model are
\begin{equation}
U(2)_{\vec{\psi}_L} \times U(2)_{\vec{u}_R} \times U(2)_{\vec{d}_R} \times U(2)_{\vec{\mathcal Q}_L} \times U(1)_{t_R} \times U(1)_{b_R}
\times U(1)_{Q_R}~.
\label{eq:flavorsymmetries}
\end{equation}
We will later use these flavor symmetries to simplify
our analysis of the fermion masses and Yukawa couplings.

\subsection{Fermion Masses and Yukawa Couplings}

The flavor properties of this model, which are the primary concern of this paper,
depend on the fermion masses and Yukawa couplings. The existence of the 
right-handed weak doublet state $Q_R$ permits the Dirac mass term
\begin{equation}
\vec{\bar{\mathcal Q
}}_L\cdot \vec{\mathcal M}\, Q_R + h.c.~,
\end{equation}
where $\vec{\mathcal M}$ is an arbitrary two-component complex mass matrix. Using the
$U(2)_{\vec{\mathcal Q}_L}$ symmetry of the quark kinetic terms, we will choose to work in a basis
in which $\vec{\mathcal M}^T=(0\ M)$ where $M$ is real and positive.  This defines what we will 
mean by $q_L$ and $Q_L$ in Table \ref{table:i} from here on.

The third-generation quark Yukawa couplings are given by
\begin{equation}
\frac{\sqrt{2}M}{v} \cdot \left(
\beta_b \bar{q}_{L} \phi  b_R + \beta_t \bar{q}_{L} \tilde{\phi} t_R
\right) + h.c.~,
\label{eq:topyukawa}
\end{equation}
where the $\beta_{t,b}$, can be chosen to be real, using the $U(1)_{t_R}\times U(1)_{b_R}$ symmetries.
The Yukawa couplings for the light two generations are given by
\begin{equation}
\frac{\sqrt{2}M}{v} \cdot \left(
\vec{\bar{\psi}}_L \phi \lambda_d \vec{d}_R + \vec{\bar{\psi}}_L \tilde{\phi} \lambda_u \vec{u}_R
\right) + h.c.~.
\end{equation}
where $\lambda_{u,d}$ are $2\times 2$ complex matrix Yukawa couplings. Neglecting the (small) mixing
of the third-generation quarks with the first two generations, the parameters $\beta_{t,b}$
and $\lambda_{u,d}$ are just equal to the corresponding parameters in the standard model,
up to the factor of $\sqrt{2}M/v$ which is included for later convenience.\footnote{While incorporating
$M$ into the Yukawa couplings is convenient for subsequent calcualtions, its presence
obscures the {\it decoupling} properties \cite{Appelquist:1974tg} of the theory in the limit
$M\to \infty$.}

Mixing of the third quark generation with the first two occurs only because all
three generations mix with the (heavy) vector quarks. Mixing between the third-generation 
quarks and the vector quarks occurs through
\begin{equation}\label{eq:4-3mix}
\frac{\sqrt{2}M}{v} \cdot \left(  \lambda'_b \bar{Q}_{L}\phi b_R+\lambda'_t \bar{Q}_{L}\tilde{\phi} t_R\right) + h.c.~,
\end{equation}
and mixing between the first- and second-generation quarks and the vector quarks occurs through the Yukawa couplings to the color-octet scalar
\begin{equation}
\frac{M}{u}\cdot \left(
\vec{\bar{\psi}}_L \cdot \vec{\alpha}\, \Phi Q_R
\right) + h.c.~.
\label{eq:octetyukawa}
\end{equation}
Here the $\lambda'_{b,t}$ are complex numbers, while $\vec{\alpha}$ is a two-component complex vector,
whose phases and orientations can be simplified using the residual flavor symmetries in a manner
that we will specify shortly. 

Note that in the limit that either $\lambda'_{b,t}\to 0$ or $\vec{\alpha} \to 0$,
third-generation quark number is conserved separately from
quark number for the light quarks,\footnote{In the limit $\lambda'_{t,b} \to 0$, top- and
bottom-quark number is conserved separately, while in the limit $\vec{\alpha} \to 0$ it is conserved
in combination with vector-quark number.} and mixing between the third generation and the first two vanishes.
Having the mixing between the heavy and light quark generations be small is therefore natural in this model.

\subsection{Quark Mass Eigenstates}
\label{sec:ckm}

The masses and Yukawa couplings above give rise to $4 \times 4$ up- and down-quark 
matrices
\begin{equation}
{\cal M}_u = M \cdot \begin{pmatrix}
\multicolumn{2}{c}{\Delta_u}
& \vec{0} & \vec{\alpha}
\\
0 & 0 & \beta_t & 0 \\
0 & 0 & \lambda'_t & 1
\end{pmatrix}~,
\ \ \ \ \ 
{\cal M}_d = M \cdot \begin{pmatrix}
\multicolumn{2}{c}{{\cal C}\Delta_d}
& \vec{0} & \vec{\alpha}
\\
0 & 0 & \beta_{b} & 0 \\
0 & 0 & \lambda'_{b} & 1
\end{pmatrix}~,
\label{eq:4x4massmatrices}
\end{equation}
in a basis where the first two components are the light-quark fields, the next is the third generation,
and the last represents the vector quarks. Here, given the small mixing of the first two generations
with the others, we use the $U(2)_{\vec{\psi}_L} \times U(2)_{\vec{u}_R} \times U(2)_{\vec{d}_R}$ symmetries to make the two-by-two matrices $\Delta_{u,d}$ real and diagonal (and given
approximately by the masses of the light quarks), while ${\mathcal C}$ is approximately
the real Cabbibo rotation matrix.

We will diagonalize these matrices, and find the corresponding
eigenstates, in the limits
\begin{equation}
\Delta_{u,d}\ \ll\ |\vec{\alpha}|,\ |\lambda'_{b,t}|,\ \beta_{b,t}\ \ll 1~.
\label{eq:conditionsi}
\end{equation}
To lowest order we find that the left-handed mass-eigenstate heavy quark field is
\begin{equation}
{\mathsf T}_{\mathsf L} = Q^u_L + \vec{\alpha}\cdot  \vec{\psi}^u_L~,
\label{eq:Tleft}
\end{equation}
while for the heavy bottom-quark we have
\begin{equation}
{\mathsf B}_{\mathsf L} = Q^d_L + \vec{\alpha} \cdot \vec{\psi}^d_L~.
\label{eq:Bleft}
\end{equation}
Note that the $\vec{\psi}^{u,d}$ correspond to the $T_3=\pm 1/2$ states in the flavor vector $\vec{\psi}$, respectively,
and that we now denote mass eigenstate fields by sanserif font.
In contrast, to lowest order, the right-handed mass eigenstate heavy fields are
\begin{equation}
{\mathsf T}_{\mathsf R} = Q^u_R + \lambda^{'*}_t t_R~,
\label{eq:Tright}
\end{equation}
and
\begin{equation}
{\mathsf B}_{\mathsf R} = Q^d_R + \lambda^{'*}_b b_R~.
\label{eq:Bright}
\end{equation}
The masses of these heavy ${\mathsf T}$ and ${\mathsf B}$ states are, including second order corrections in the Yukawa couplings,
\begin{eqnarray}
M_{\mathsf T} &=& M + \dfrac12 \left( |\lambda^{'}_t|^2 + |\vec{\alpha}|^2 \right) M \\
M_{\mathsf B} &=& M + \dfrac12 \left( |\lambda^{'}_b|^2 + |\vec{\alpha}|^2 \right) M
\end{eqnarray}
so, given the limit in Eq. (\ref{eq:conditionsi}),  they are both of order $M$.

By ``integrating out" the heavy ${\mathsf T}$ and ${\mathsf B}$ fields, we find that the effective $3 \times 3$ mass 
matrices for the up- and down-quarks have the form
\begin{equation}
{\mathsf M}_u = M \cdot \begin{pmatrix}
\multicolumn{2}{c}{\Delta_u} & -\lambda'_t \vec{\alpha}  \\
\multicolumn{2}{c}{0} & \beta_t
\end{pmatrix}~,\ \ \ \ \ 
{\mathsf M}_d = M \cdot \begin{pmatrix}
\multicolumn{2}{c}{{\cal C}\Delta_d} & -\lambda'_b \vec{\alpha}  \\
\multicolumn{2}{c}{0} & \beta_b
\end{pmatrix}~.
\label{eq:3x3massmatrices}
\end{equation}
Since we know that $m_t\gg m_{u,c}$ and $m_b\gg m_{d,s}$, we can further assume
that
\begin{equation}
\left| \frac{\lambda'_t \vec{\alpha}}{\beta_t}\right|~,\,
\left| \frac{\lambda'_b \vec{\alpha}}{\beta_b}\right| \ll 1~,
\end{equation}
and we find that the left-handed third-generation mass eigenstates are approximately given by
\begin{equation}
{\mathsf t}_{\mathsf L} =  q^t_{L} - \frac{\lambda'_t \vec{\alpha}}{\beta_t} \cdot \vec{\psi}^u_L~,
\label{eq:fcnctop}
\end{equation}
\begin{equation}
{\mathsf b}_{\mathsf L} =  q^b_{L} - \frac{\lambda'_b \vec{\alpha}}{\beta_b} \cdot \vec{\psi}^d_L~.
\label{eq:fcncbottom}
\end{equation}

The $3 \times 3$ rotation matrix for up-type left-handed quarks\footnote{In particular, the matrix $U_L$ is defined so that $U^\dagger_L {\mathsf M}_U {\mathsf M}^\dagger_U U_L$ is diagonal, real, and
positive -- and similarly for the matrix $D_L$ with respect to the down-quark mass matrix ${\mathsf M}_D$.}  is:
\begin{equation}\label{matrix:UL}
U_L =\begin{pmatrix} 1 & 0 & -\frac{\alpha_1 \lambda^{'}_t}{\beta_t} \\  0 & 1 & -\frac{\alpha_2 \lambda^{'}_t}{\beta_t}  \\   \left(\frac{\alpha_1 \lambda^{'}_t}{\beta_t}\right)^* & \left(\frac{\alpha_2 \lambda^{'}_t}{\beta_t}\right)^* & 1 \end{pmatrix} 
\end{equation}
while the rotation matrix for down-type left-handed quarks is:
\begin{equation}\label{matrix:DL}
D_L =\begin{pmatrix} V_{ud} & V_{us} & -\frac{\alpha_1 \lambda^{'}_b}{\beta_b} \\  V_{cd} & V_{cs} & -\frac{\alpha_2 \lambda^{'}_b}{\beta_b}  \\   \left(\frac{\lambda^{'}_b}{\beta_b}\right)^* (V_{ud}\alpha^*_1 +V_{cd}\alpha^*_2) & \left(\frac{\lambda^{'}_b}{\beta_b}\right)^* (V_{us}\alpha^*_1 +V_{cs}\alpha^*_2)  & 1 \end{pmatrix} 
\end{equation}
At leading order in $\alpha$, $\lambda'_t$, and $\lambda'_b$ the CKM mixing matrix $V_{CKM}= U^\dag_L D_L$ is therefore
\begin{equation}\label{matrix:ckm}
V_{CKM}=\begin{pmatrix} V_{ud} & V_{us} & \alpha_1 d \\  V_{cd} & V_{cs} & \alpha_2 d \\-(V_{ud}\alpha^*_1+V_{cd}\alpha^*_2)d^* & -(V_{us}\alpha^*_1+V_{cs}\alpha^*_2)d^* & 1 \end{pmatrix} 
\end{equation}
\noindent
where
\begin{equation}
d\equiv\frac{\lambda'_t}{\beta_t}-\frac{\lambda'_b}{\beta_b}~,
\label{eq:reali}
\end{equation}
and the upper $2\times 2$ block corresponds to the previously-mentioned matrix ${\cal C}$. 

Using the third-generation quark number symmetry\footnote{Rotations using this symmetry are made once $U(1)_{\vec{q}_L}$ and $U(1)_{t_R}\times U(1)_{b_R}$ are used to make $M,\beta_t, \beta_b$ all real.}, we can choose
\begin{equation}
\frac{{\rm Im}\, \lambda'_t}{\beta_t} =  \frac{{\rm Im}\, \lambda'_b}{\beta_b} ~,
\label{eq:reallii}
\end{equation}
which ensures that the combination $d$, above, is real.  We can also use the quark number symmetry for the first two generations\footnote{This freedom remains after $U(2)_{\vec{\psi}_L} \times U(2)_{\vec{u}_R} \times U(2)_{\vec{d}_R}$ is used to put the entries of ${\cal M}_u$ and ${\cal M}_d$ into the form shown in Eq. (\ref{eq:4x4massmatrices}).} to adjust the phase
of
\begin{equation}
\vec{\alpha} 
\equiv \begin{pmatrix}
\alpha_1 \\
\alpha_2
\end{pmatrix}~,
\end{equation}
so that $\alpha_2$ is real. In this basis, the CKM matrix of Eq. (\ref{matrix:ckm}) has the conventional
form and the combination of parameters can be compared to the Wolfenstein parameterization
\begin{equation}\label{eq:wolf}
\begin{pmatrix}
1-\frac{\lambda^2}{2} & \lambda & A \lambda^3(\rho-i\eta)\\
-\lambda & 1-\frac{\lambda^2}{2} & A \lambda^2 \\
A \lambda^3(1-\rho-i\eta) & -A\lambda^2 & 1
\end{pmatrix}~,
\end{equation}
(which is good up to corrections ${\cal O}(\lambda^4)$). The measured values of the parameters are \cite{Beringer:1900zz}
\begin{align}
\lambda=0.22535 &\pm 0.00065 \\
A=0.817 & \pm 0.015\\
\rho = 0.140 & \pm 0.018 \\
\eta = 0.357 & \pm 0.014~.
\end{align}

Comparing Eq. (\ref{matrix:ckm}) and (\ref{eq:wolf}), we find
that this model consistently reproduces the observed pattern of
quark mixing if $\alpha_1={\cal O}(\lambda^2)$, $\alpha_2 = {\cal O}(\lambda^3)$, and both $\lambda'_t/\beta_t$ and $\lambda'_b/\beta_b$ are of order 1. Note that, from Eq. (\ref{eq:topyukawa}), we can estimate that
\begin{equation}
\beta_{b,t} \simeq \frac{m_{b,t}}{M}~,
\end{equation}
and, therefore, $\beta_{b,t}$ and (hence $\lambda'_{t,b}$)
can be made arbitrarily small by
taking the vector quark Dirac mass $M$ to be large enough.\footnote{Note, however, that we must
ensure that the color-octet Yukawa coupling of Eq. (\ref{eq:octetyukawa}) remains
perturbative, $M|\alpha_{1,2}|/u \le 4\pi$.}

Summing up, we have found that the CKM matrix is correctly reproduced by:
\begin{align}
\begin{split}\label{est-ckm}
& V_{ub}=\alpha_1 d=\alpha_1 \left(\frac{\lambda'_t}{\beta_t}-\frac{\lambda'_b}{\beta_b}\right)=A\lambda^3(\rho-i\eta)=0.00131-i0.00334\\
& V_{cb}=\alpha_2 d=\alpha_2 \left(\frac{\lambda'_t}{\beta_t}-\frac{\lambda'_b}{\beta_b}\right)=A\lambda^2=0.0415~,
\end{split}
\end{align}
where both $d$ and $\alpha_2$ are real, and where $\alpha_2$ is ${\cal O}(\lambda^2)$ while $\alpha_1$ is ${\cal O}(\lambda^3)$.

 Reproducing CKM mixing, however, is not sufficient
to insure that the model is consistent with observed flavor physics since the model incoroporates
non-standard interactions. In the next section we review the additional contributions to
various flavor-changing neutral current processes and consider the resulting bounds on the parameters.

\section{Constraints from Flavor-Changing Processes}

New contributions to flavor-changing neutral currents (FCNC) arise in our model from the mixing of the ordinary fermions with the new weak vector femion states, and through the couplings of the colorons to fermions.  We find that data on $b \to s\gamma$ and $\Delta F = 2$ meson mixing processes place bounds on the model parameters but leave substantial regions of allowed parameter space.

\subsection{Limits from $b \to s\gamma$}

\begin{figure}
\includegraphics[width=5in]{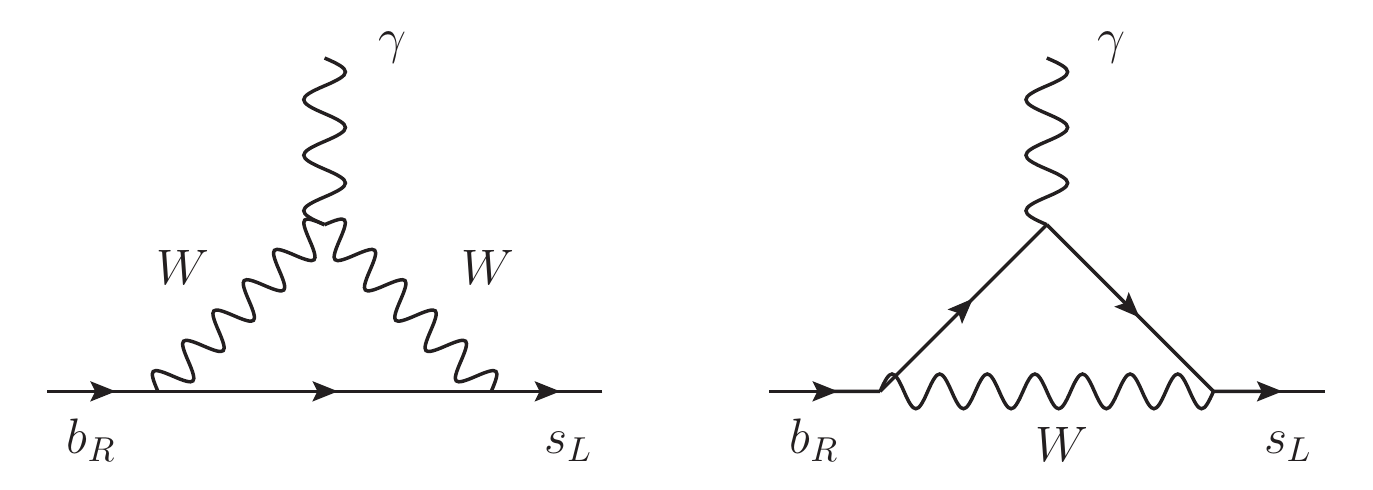}
\caption{Generic form of one-loop contributions to $b\to s \gamma$.}
\label{fig:b-s-g}
\end{figure}

The top-coloron model includes additional weak vector fermions and therefore, through the mixing terms
controlled by $\lambda'_{t,b}$ as shown in Eqs. (\ref{eq:Tright}) and (\ref{eq:Bright}), induced 
electroweak interactions of the right-handed quarks. Such interactions give rise to enhanced
contributions to the process $b \to s\gamma$. The generic form of the diagrams that contribute to this process are sketched in Fig. \ref{fig:b-s-g}. Some contributions due to new physics arise from the induced right-handed
couplings, $W t_R b_R$; others result when a heavy vector quark is exchanged
in the loop. We consider each kind of contribution below in turn.

We begin by computing the right-handed couplings of the 
mass-eigenstate fermion fields to the $W$-boson, 
which are related to the left- and right-handed matrices that diagonalize the $4\times 4$ quark 
mass matrices in Eq. (\ref{eq:4x4massmatrices}). To leading order in the down quark sector
we find
\begin{equation}\label{matrix:DL4}
{\mathcal D}_L = \begin{pmatrix} V_{ud} & V_{us} & 0 & \alpha_1  \\  V_{cd} & V_{cs} & 0 & \alpha_2 \\
0 & 0 & 1 & 0 \\-(V_{ud}\alpha^*_1+V_{cd}\alpha^*_2) & -(V_{us}\alpha^*_1+V_{cs}\alpha^*_2) & 0 &1 \end{pmatrix}
\end{equation}
and 
\begin{equation}\label{matrix:DR4}
{\mathcal D}_R=\begin{pmatrix} 1 & 0 & 0 & 0  \\  0 & 1 & 0 & 0  \\
0 & 0 & 1 & \lambda^{'*}_b \\
0 & 0 & -\lambda^{'}_b &1 \end{pmatrix}~.
\end{equation}
\noindent
There are similar expressions in the up-quark sector for rotation matrices
 ${\mathcal U}_{L,R}$, which are obtained from the ${\mathcal D}_{L,R}$ by setting $\lambda'_b \to \lambda'_t$ and 
$V_{ud},V_{cs} \to 1$ and $V_{us},V_{cd} \to 0$.

\begin{figure}[ht]
\begin{minipage}[b]{0.35\linewidth}
\centering
\includegraphics[width=0.8\textwidth]{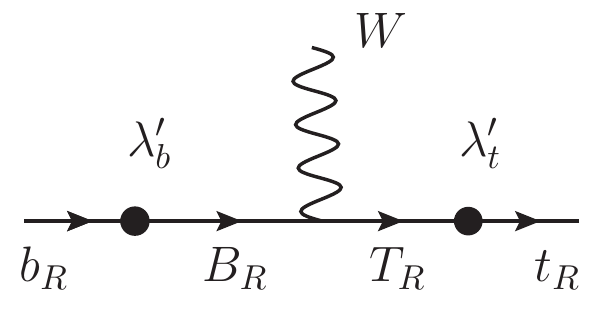}
\caption{Origin of the effective $W t_R b_R$ vertex in our coloron model.}
\label{fig:RH-Wtb}
\end{minipage}
\hspace{0.5cm}
\begin{minipage}[b]{0.55\linewidth}
\centering
\includegraphics[width=0.8\textwidth]{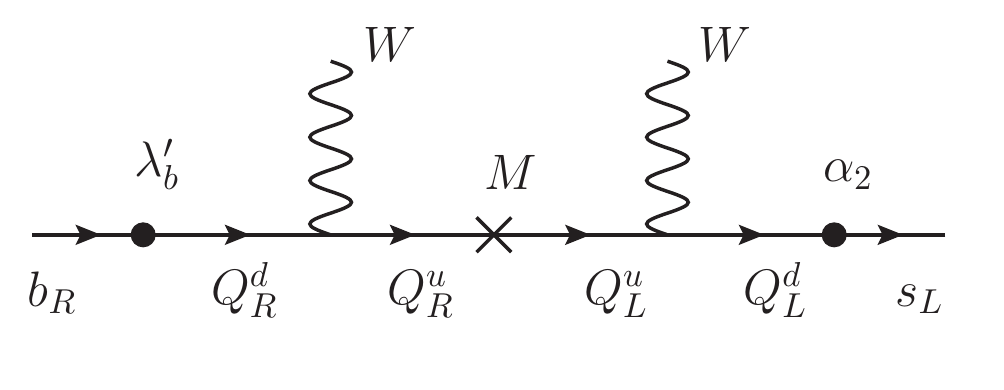}
\caption{Combination of effective $W b_R Q^u_R$ and $W Q^u_L s_L$ vertices contributing to $b\to s \gamma$ in our coloron model}
\label{fig:bsgamma}
\end{minipage}
\end{figure}

The presence of an effective $W t_R b_R$ vertex has been discussed in ref. \cite{Grzadkowski:2008mf}.   
The sketch in Fig.~\ref{fig:RH-Wtb} shows that the size of this coupling is
given by
\begin{equation}
\frac{ig}{\sqrt{2}} \cdot ({\mathcal D}^\dagger_R)_{34}({\mathcal U}_R)_{43} = \frac{ig}{\sqrt{2}} \cdot \left(\lambda'_t \lambda^{'*}_b \right)
\end{equation}
in this model.
Given the experimental limits on the rates of $b \to s \gamma$ arising from induced $W t_R b_R$ couplings that are quoted in ref. \cite{Grzadkowski:2008mf}, we conclude that
\begin{equation}\label{eq:vR-lim}
-0.0007 \leq \rm Re [\lambda'_t \lambda^{'*}_b] \leq 0.0025
\end{equation}
Note that if either $\lambda'_b=0$ or $\lambda'_t=0$, this limit is automatically satisfied without providing additional information about the size of the other coupling.


There is also an additional contribution to $b\to s \gamma$ due to exchange of the heavy quarks $Q$, which arises from the fact that $Q_R^d$ mixes with $b_R$ as shown in Eq. (\ref{eq:Bright})  and $Q^u_L$ mixes with $\vec{\psi}^u_L$ as shown in Eq. (\ref{eq:Bleft}). The combination of an effective $W b_R Q^u_R$ vertex with an effective $W Q^u_L s_L$ vertex, as sketched in Fig. \ref{fig:bsgamma}, yields a new contribution to the $b\to s \gamma$ decay.  

To evaluate the size of the contribution from $Q$ exchange, we start by writing down the effective Hamiltonian for $b\to s \gamma$ decay \cite{Buras}
\begin{equation}\label{eq:Heff-bsgamma}
H_{eff}=-\frac{4G_F}{\sqrt{2}} V^{*}_{ts}V_{tb}\left[C_7 O_7+ C^{'}_7 O^{'}_7 \right]\,,
\end{equation}
where $O_7\equiv\frac{e m_b}{16\pi^2} \bar{s} \sigma^{\mu\nu} \frac{1+\gamma_5}{2}b F_{\mu\nu}$ and $O^{'}_7\equiv\frac{e m_b}{16\pi^2} \bar{s} \sigma^{\mu\nu} \frac{1-\gamma_5}{2}b F_{\mu\nu}$.
\noindent
The Wilson coefficient $C_7$ in our model is obtained by evaluating how the new physics shown in the Fig. \ref{fig:bsgamma} contributes to the  
 loop diagrams of Fig. \ref{fig:b-s-g}.  The result is
\[
C_7= ({\mathcal D}^{\dag}_R)_{34}({\mathcal D}_L)_{42} \frac{M}{m_b} \frac{\tilde{F}(x)}{V^{*}_{ts}V_{tb}}\,,
\]
where 
\begin{equation}\label{eq:fun-bsgamma}
\tilde{F}(x)=\frac{-20+31x-5 x^2}{12(x-1)^2}+\frac{x(2-3x)}{2(x-1)^3}\ln x
\end{equation}
is the loop function of $x\equiv M^2/M^2_w$ calculated in \cite{Fujikawa:94,Cho:93}, in the context of Left-Right symmetric models. We find that $\tilde{F}$ takes values between -0.46 and -0.42 for $x$ lying in the range $[(1 \text{TeV}/M_w)^2, \infty]$. The coefficient
\[
({\mathcal D}^{\dag}_R)_{34}({\mathcal D}_L)_{42}= \lambda^{'*}_b\alpha^{*}_2+O(\lambda^4)
\]
is the product of the coefficients of the effective $\bar{b}_R W Q^u_R$ vertex, generated by the mixing
 $\bar{B}_R\to \bar{b}_R$, and of the vertex $\bar{Q}^u_L W s_L$, coming from $B_L\to s_L$. We have thus
\begin{equation}\label{eq:C7}
C_7=\lambda^{'*}_b\alpha^{*}_2\frac{M}{m_b} \frac{\tilde{F}(x)}{V^{*}_{ts}V_{tb}}=\frac{\lambda^{'*}_b\alpha^{*}_2}{\beta_b}\frac{1}{V^{*}_{ts}V_{tb}}\tilde{F}(x)=-\frac{\lambda^{'*}_b/\beta_b}{\lambda^{'*}_t/\beta_t-\lambda^{'*}_b/\beta_b}\tilde{F}(x)\,,
\end{equation}
where in the last equality we have used the identity in Eq.~(\ref{est-ckm}).
The contribution to $O^{'}_7$ is suppressed by mass insertions in the external quark legs and, thus, leads to weaker constraints than those from the contribution to $O_7$.

The experimental measurement of the $b\to s\gamma$ branching ratio \cite{exp-bsgamma} leads to the following $95\%$ C.L. limits on the coloron model contribution to $O_7$ at the TeV scale (Appx.~\ref{App_C7_bound}):
\begin{equation}\label{eq:C7-bound}
-0.093<\rm Re [C_7 (1\ \text{TeV})]<0.023\,.
\end{equation}
\noindent 
We can use this both to constrain the relative sizes of the top-sector and bottom-sector couplings and also to place a bound on the absolute size of the bottom-sector couplings. 

\begin{figure}
\centering
\includegraphics[width=0.5\textwidth]{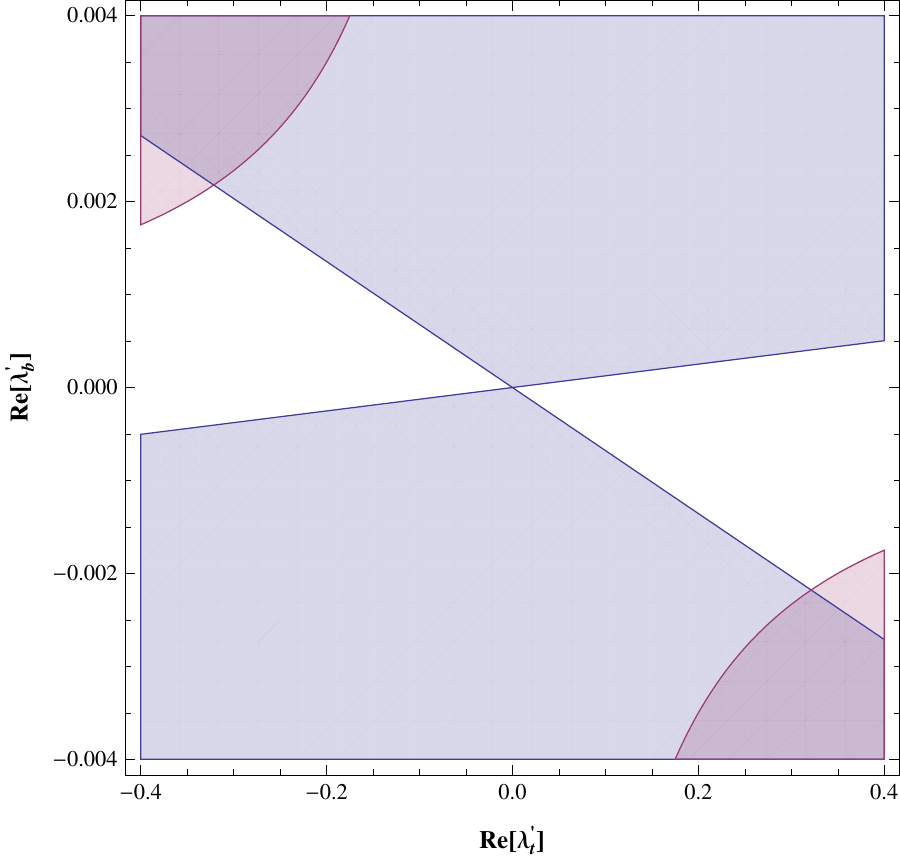} 
\caption{Region (unshaded) of the ($\rm Re [\lambda^{'}_t]$, $\rm Re [\lambda^{'}_b]$) plane that is allowed by $b\to s\gamma$. The blue region (upper right and lower left) is excluded by the bound in (\ref{eq:ratio-C7-bound}), coming from the constraint on $\rm Re [C_7(1 \text{TeV})]$; the small pink regions (upper left and lower right corners) are excluded by the limit (\ref{eq:vR-lim}) on $\lambda'_t \lambda^{'*}_b$.  As discussed in the text, we have assumed $\rm Im [\lambda^{'}_b]=0$ and $\beta_t/\beta_b=m_t/m_b$.}
\label{fig:Excl-lambda-tb}
\end{figure}

To study the relative sizes of $\lambda'_t$ and $\lambda'_b$, we recall the relation in Eq. (\ref{eq:reallii}).  This implies both that the denominator of the right-hand-side of Eq. (\ref{eq:C7}) is real, and also that
${\rm Im} \lambda'_b = (m_b/m_t)\, {\rm Im} \lambda'_t$. If we neglect ${\rm Im}\lambda'_b$, 
the bound in (\ref{eq:C7-bound}) yields the following constraints on the ratio $\rm Re[\lambda^{'}_t]/\rm Re[\lambda^{'}_b]$:
\begin{equation}\label{eq:ratio-C7-bound}
\frac{\rm Re[\lambda^{'}_t]}{\rm Re[\lambda^{'}_b]}< -3.9 \frac{\beta_t}{\beta_b} \qquad\qquad \frac{\rm Re[\lambda^{'}_t]}{\rm Re[\lambda^{'}_b]}> 21 \frac{\beta_t}{\beta_b}\,,
\end{equation}
which excludes the case $\rm Re [\lambda^{'}_t]=0$.
Fig.~\ref{fig:Excl-lambda-tb} shows (unshaded) the  region in the ($\rm Re [\lambda^{'}_t]$, $\rm Re [\lambda^{'}_b]$) plane that is allowed by $b\to s\gamma$ after one applies the (pink, less restrictive) limits from the effective $Wt_Rb_R$ vertex (\ref{eq:vR-lim}) and the (blue, more restrictive) limits from $Q$ exchange (\ref{eq:ratio-C7-bound}). 

We can also find a limit on the size of the $b$ couplings by inserting the observed values of $V_{ts}$ and $V_{tb}$ into the middle term of Eq. (\ref{eq:C7}) and comparing it to Eq. (\ref{eq:C7-bound}).  The result is
\begin{equation}\label{eq:alpha2-C7-bound}
-0.0085<\frac{\alpha_2 Re[\lambda^{'}_b]}{\beta_b}<0.0021~.
\end{equation}
\noindent
We conclude that (in the limit where $\rm Im[\lambda^{'}_b]\ll \rm Re[\lambda^{'}_b]$) the maximum magnitude of $\bigg|\frac{\alpha_2 \lambda^{'*}_b}{\beta_b}\bigg|$ is 0.0085 -- a value that will be useful later on.

\subsection{Limits from $\Delta F = 2$ FCNC}

The couplings of the massive
coloron to fermions (see Eqs. (\ref{eq:coloron}) and (\ref{eq:coloroncurrent}))
are not flavor-universal and, therefore, coloron exchange will generate flavor-changing
neutral currents. We begin by considering flavor mixing in the $B$-meson system, and then turn
to the stronger constraints arising from $D$ and $K$ meson mixing.

\subsubsection{Mixing involving light quarks and heavy quarks}

For B-meson mixing, the $b$-quark mass eigenstates are approximately gauge-eigenstates of $SU(3)_1$,
while the mass eigenstates for the light quarks (s or d) are approximateliy $SU(3)_2$ gauge-eigenstates.  
Therefore, neutral meson mixing arises due to the presence, in the quark mass eigenstates 
$\mathsf{t_L}$ and $\mathsf{b_L}$, of mixing between the top-bottom doublet $q_L$ and light $\psi_L$ states, as shown in Eq. (\ref{eq:fcncbottom}).

\begin{figure}
 \includegraphics[width=0.8\textwidth]{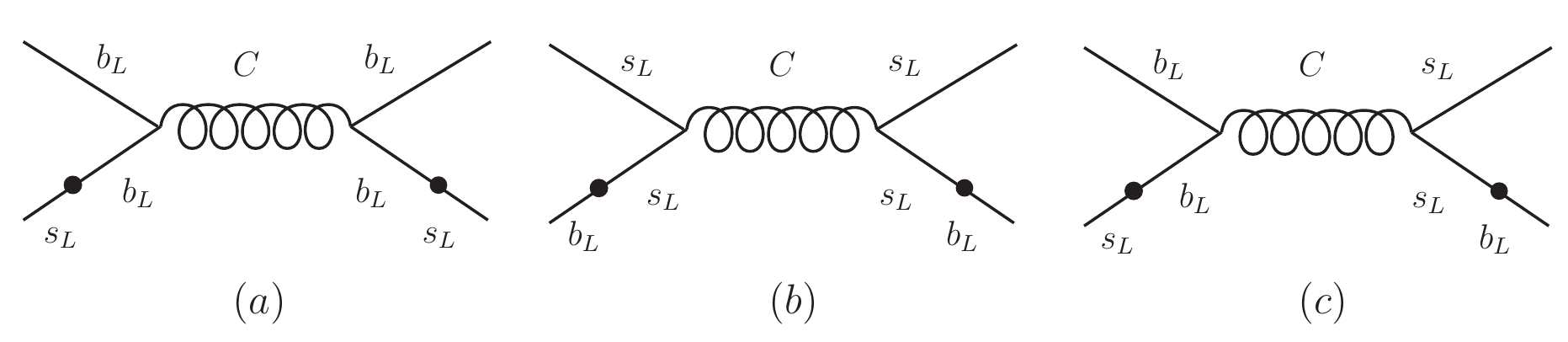}
\caption{Sketch of three contributions to $B_s$ meson mixing with different dependences on the coloron mixing angle $\omega$.}
\label{fig:bsbs-diag}
\end{figure}

The 4-fermion operator $\left(\bar{b}_L\gamma^\mu s_L\right)^2$ receives three types of contribution from coloron exchange, depending on which quarks are directly interacting with the coloron:
\begin{itemize}
\item $g^2_S\left(\cot\omega J^{a\mu}_{bL}\right)^2 (D_L)^2_{32}$, as in Fig. \ref{fig:bsbs-diag}(a)
\item $g^2_S\left(-\tan\omega J^{a\mu}_{sL}\right)^2 (D^\dag_L)^2_{32}$, as in Fig. \ref{fig:bsbs-diag}(b)
\item $2g^2_S\left(-\tan\omega J^{a\mu}_{sL}\right)\left(\cot\omega J^{a\mu}_{bL}\right) (D_L)_{32}(D^\dag_L)_{32}$, as in Fig. \ref{fig:bsbs-diag}(c)
\end{itemize}
where $D_L$ is the matrix in (\ref{matrix:DL}). As shown in Sec. \ref{sec:ckm}, to an accuracy of $O(\lambda^4)$ we can explicitly write:
\begin{align}
\begin{split}
&(D_L)_{32}=\left(\frac{\lambda'_b}{\beta_b}\right)^* \left(V_{us}\alpha^*_1+V_{cs}\alpha^*_2 \right)=\left(\frac{\lambda'_b \alpha_2}{\beta_b}\right)^*\\
& (D^\dag_L)_{32}=-\left(\frac{\lambda'_b \alpha_2}{\beta_b}\right)^*
\end{split}
\end{align}

Summing over the three contributions, we get the Wilson coefficient\footnote{Here we follow the
notation of \cite{Bona:2007vi}.} of the operator $(\bar{b}_L \gamma^\mu s_L)^2$ for $B_s$-mixing:
\begin{align}
\begin{split}
C^1_{Bs}& = \frac{1}{6}\frac{g^2_S}{M^2_C}\left[\cot^2\omega\left(\left(\alpha_2\frac{\lambda'_b}{\beta_b}\right)^*\right)^2+\tan^2\omega\left(\left(\alpha_2\frac{\lambda'_b}{\beta_b}\right)^*\right)^2+2\cot\omega\tan\omega \left(\alpha_2\frac{\lambda'_b}{\beta_b}\right)^*\left(\alpha_2\frac{\lambda'_b}{\beta_b}\right)^*\right]\\
&=\frac{1}{6}\frac{g^2_S}{M^2_C}\left(\left(\alpha_2\frac{\lambda'_b}{\beta_b}\right)^*\right)^2\left[\cot\omega+\tan\omega\right]^2
\end{split}
\end{align} 
%
%
The coefficient $C^1_{Bd}$ for $B_d$ meson mixing is analagous, but  depends instead on
\begin{align}
\begin{split}
&(D_L)_{31}=\left(\frac{\lambda'_b}{\beta_b}\right)^* \left(\alpha^*_1-\lambda\alpha^*_2 \right) + O(\lambda^4)\\
& (D^\dag_L)_{31}=-\left(\alpha_1\frac{\lambda'_b}{\beta_b}\right)^*
\end{split}
\end{align}

The UTFit collaboration has provided a valuable summary of limits on operators producing flavor-changing neutral currents; the original review is \cite{Bona:2007vi} and the most recent update is in \cite{Bona:Vietnam}.  
Applying their results to our case we find the limits
\begin{equation}
M_C > 175 \ g_S \bigg|\dfrac{\lambda^{'}_b}{\beta_b}\bigg|\cdot \bigg|(\alpha^*_1-\lambda\alpha^*_2)\cot\omega +\alpha^*_1\tan\omega \bigg|\ {\rm TeV}~,
\end{equation}
from constraints on additional contributions to $B_d$-mixing, and 
\begin{equation}
M_C > 41 \ g_S \bigg|\dfrac{\alpha_2 \lambda'_b}{\beta_b}\bigg| \cdot \bigg|\cot\omega+\tan\omega \bigg|\ {\rm  TeV}
\label{eq:mcgreater}
\end{equation}
from $B_s$-mixing. 

\begin{figure}
\includegraphics[width=0.6\textwidth]{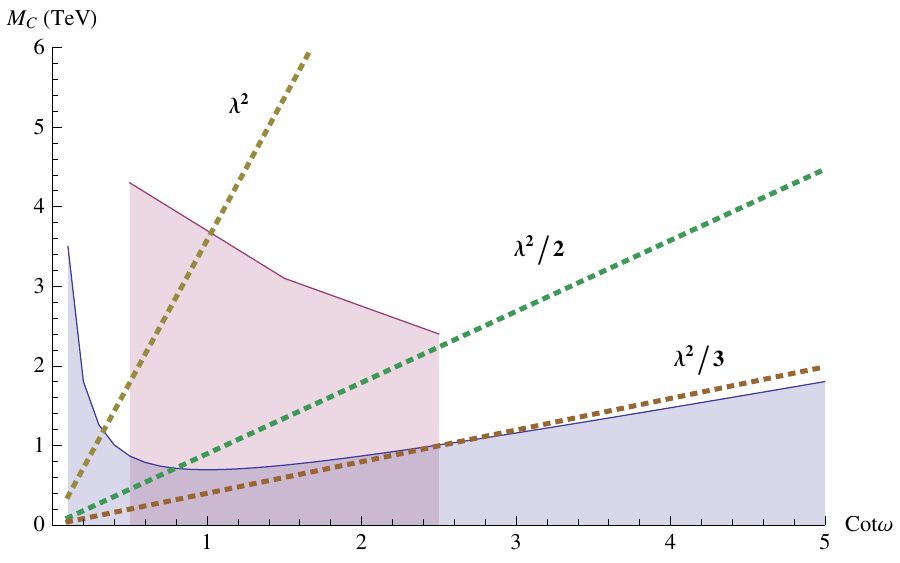}
\caption{Left:  Exclusion regions on the plane $(\cot\omega, M_C)$ from the ATLAS search for dijet resonances (pink region, beneath the short upper curve) and from $B_s$ mixing, as in Eq. (\ref{eq:BBlimit}), assuming that $\bigg|\frac{\alpha_2 \lambda^{'}_b}{\beta_b}\bigg|$ takes on the maximum value allowed by $b\to s \gamma$ (blue region, beneath the long lower curve). In addition, the bound on contributions to $\rm Im [C^1_K]$ in Eq. (\ref{eq:kk-limit}) excludes the region below the dashed line whose label matches the value of $|\alpha_2|$; a larger value yields a stronger bound. }
\label{fig:coupling-plane-sketch}
\end{figure}

Note that these bounds are proportional to $\lambda'_b$.  Recall that to reproduce
CKM mixing, we need only satify
\[
\dfrac{\alpha_2 \lambda'_t}{\beta_t}-\dfrac{\alpha_2 \lambda'_b}{\beta_b} =A\lambda^2
\]
which leaves open the possibility that $\lambda'_b=0$, in which case the model would not be constrained by $B$ mixing or by $b\to s\gamma$ (as discussed previously). 
For non-zero $\lambda'_b$, we expect the 
constraint from contributions to $B_s$-mixing to be stronger than that from $B_d$, since ${\cal O}(\alpha_1) \simeq \lambda {\cal O}(\alpha_2)$.
If we take the maximum value $|\alpha_2 \lambda'_b/\beta_b|=0.0085$ from Eq. (\ref{eq:alpha2-C7-bound}), 
we get the following bound on the coloron mass from the $B_s B_s$ mixing: 
\begin{equation}\label{eq:BBlimit}
M_C> 0.35 \ g_S \ |\cot\omega+\tan\omega| \ \text{TeV}
\end{equation}
The corresponding exclusion region is shown in Fig. \ref{fig:coupling-plane-sketch}.  As indicated by Eq. (\ref{eq:mcgreater}), the size of the upper bound would scale linearly with $|\alpha_2 \lambda'_b/\beta_b|$ as one reduced this ratio below its maximum allowed value.

\subsubsection{Mixing involving light quarks only}

In $K$- and $D$-mexon mixing processes, all of the quarks are light and, to leading order, 
transform under the same color group.  Coloron exchange
contributions to neutral meson mixing arise from the mixing of the
left-handed components of the light-quarks with the heavy quarks,
$Q_L$ and $\psi_L$ states, as shown in equations (\ref{eq:Tleft}) and (\ref{eq:Bleft}).

Following the same reasoning as above, we find that this results in the operators
\begin{eqnarray}
& &\frac{1}{6}\frac{g^2_S}{M^2_C} (({\mathcal D}^\dag_L)_{14}({\mathcal D}_L)_{42})^2 \cot^2\omega (\bar{d}_L \gamma^\mu s_L)^2\\
& &\frac{1}{6}\frac{g^2_S}{M^2_C} (({\mathcal U}^\dag_L)_{14}({\mathcal U}_L)_{42})^2 \cot^2\omega (\bar{u}_L \gamma^\mu c_L)^2
\end{eqnarray}
where ${\mathcal D}_L$ is the rotation matrix in Eq. ({\ref{matrix:DL4}}) and ${\mathcal U}_L$ is the corresponding
matrix for up-quarks (obtained from ${\mathcal D}_L$ by setting $\lambda$ to zero). We find
\[
({\mathcal D}^\dag_L)_{14}({\mathcal D}_L)_{42}=\alpha_1\alpha^*_2 +\lambda (\alpha_1\alpha^*_1 - \alpha_2\alpha^*_2) -\lambda^2(\alpha_2
\alpha^*_1 +\alpha_1\alpha^*_2) - \frac{\lambda^3}{2}(\alpha_1\alpha^*_1 -\alpha_2\alpha^*_2) + O(\lambda^4) \,. \]
Inserting the known value of the ratio $\alpha_1/\alpha_2$ from Eq. (\ref{est-ckm}), we obtain
\begin{eqnarray}
& &({\mathcal D}^\dag_L)_{14}({\mathcal D}_L)_{42}=-|\alpha_2|^2(0.190+ i 0.0804)+ O(\lambda^4)~,\\
& &({\mathcal U}^\dag_L)_{14}({\mathcal U}_L)_{42}=+|\alpha_2|^2(0.0316 - i 0.0804)~.
\end{eqnarray}

We again draw on the UTFit data  \cite{Bona:2007vi} and find that their limits translate as follows:
\begin{itemize}
\item The constraint from additional contributions to CP-violation in $K$-meson mixing (based on $\rm Im[C^1_K]$) implies that  
\begin{equation}\label{eq:kk-limit}
M_C > 1.4 \cdot 10^3 \ g_S\ |\alpha_2|^2 \cot\omega \ \text{TeV}~.
\end{equation}
\item The constraint from contributions to $K$-mexon mixing (based on $\rm Re[C^1_K$]) implies that 
\begin{equation}
M_C > 82 \ g_S\ |\alpha_2|^2 \cot\omega\ \text{TeV}~.
\end{equation}

\item The limit from $D$-meson mixing (based on $C^1_D$) implies that %
\begin{equation}
M_C > 39 \ g_S\ |\alpha_2|^2 \cot\omega\  \text{TeV}~.
\end{equation}

\end{itemize}
We see that the strongest constraint comes from the limit on CP-violating contributions to $K$-meson mixing in 
Eq. (\ref{eq:kk-limit}), and this constraint is plotted (for various values of $|\alpha_2|={\cal O}(\lambda^2)$) in Fig. \ref{fig:coupling-plane-sketch}.\footnote{The corrections to the process $Z \to b\bar{b}$ yield constraints weaker than those considered above. In particular, the tree-level contribution arises from a process similar to that shown in Fig. \ref{fig:RH-Wtb}, and is of order ${\cal O}(\lambda'_b)^2$. It is therefore negligible due to the constraints from $b \to s \gamma$ displayed in Fig. \ref{fig:Excl-lambda-tb}. At one-loop order, vertex corrections are suppressed both by loop factors as well as mixing of the sort illustrated in Fig. \ref{fig:bsbs-diag}, and hence are even smaller.}

\section{Collider mass limits on Colorons}

The LHC experiment CMS \cite{cms-pas-exo-12-016} has set a limit on the mass of a flavor-universal coloron or axigluon (one that couples to all six ordinary quark flavors in the same way) in a recent paper, based on the data collected at a center-of-mass energy of 8 TeV. The flavor-universal axigluon or coloron model gives the same cross-section prediction as our more general coloron model when $\cot\omega$ takes on the value $1$ (or, equivalently when $\cos\omega = 1/\sqrt{2}$).  So from this CMS paper we can see immediately that the limit they set on colorons in this model for $\cot\omega = 1$ is about 3.3 TeV.  

ATLAS \cite{atlas-conf-2012-148} has also presented limits on new resonances decaying to dijets, based on 8 TeV data.  In that paper they did not happen to show the theoretical prediction for a flavor-universal axigluon or coloron in their plots or quote a limit on such a state.  However, since they did provide a plot showing how they set limits on hypothetical narrow particles, using simplified Gaussian models, we can just overlay our model's predicted cross-section curve on that plot and see the approximate limit from the new data ourselves.

\begin{figure}
\includegraphics[width=.5\textwidth]{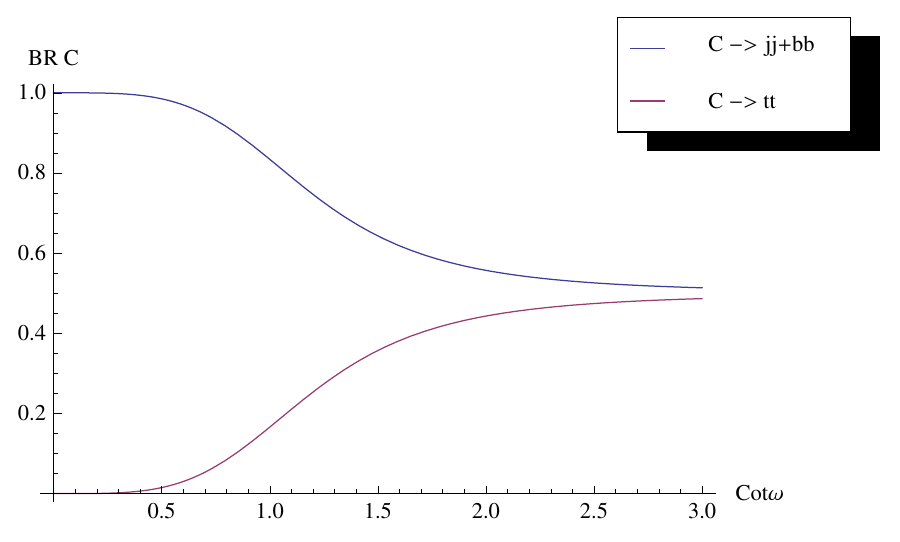}
\caption{Branching ratios for coloron decay as a function of $\cot\omega$, for $M_C=1$ TeV.  The blue upper curve is for dijet plus $b\bar{b}$ decays; the purple lower curve is for decays to top-quark pairs.}
\label{BR-C}
\end{figure}

\begin{figure}
\includegraphics[width=.45\textwidth]{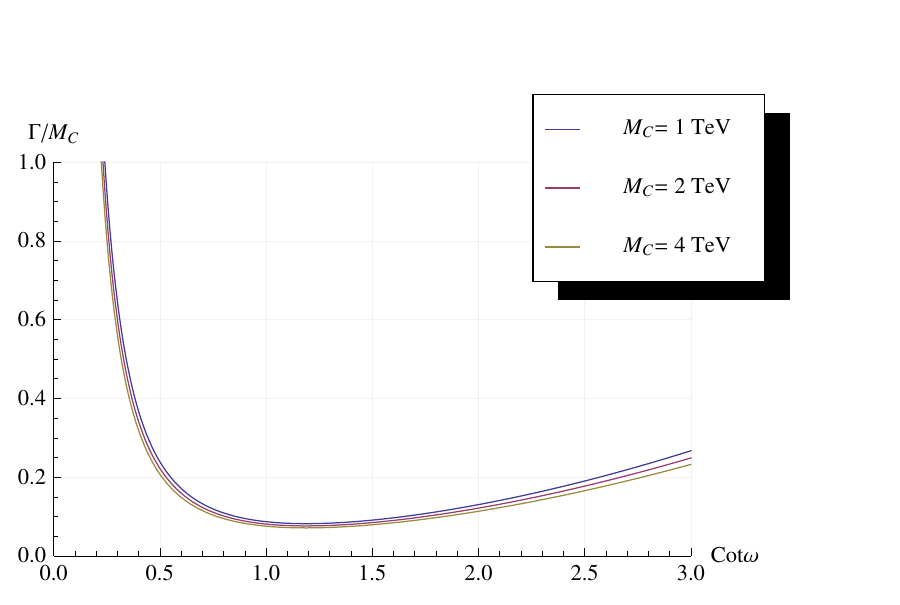}
\includegraphics[width=.45\textwidth]{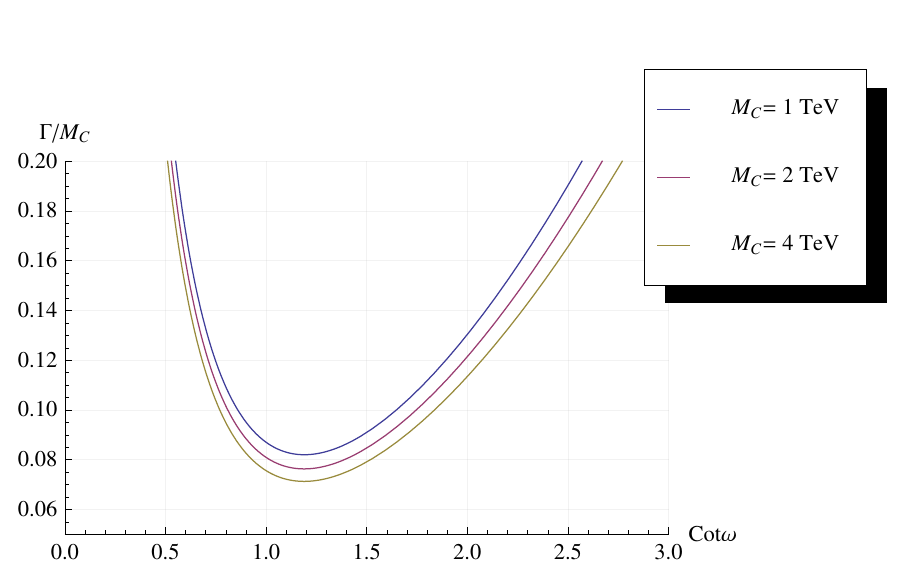}
\caption{Coloron decay width over coloron mass, shown as a function of $\cot\omega$, for $M_C=1,2,4$ TeV (respectively, the upper, middle, and lower curves). The right-hand plot is a close-up of the low $\Gamma/M$ region of the left-hand plot.}
\label{width-C}
\end{figure}

In order to understand which data set can probe which values of the coloron mixing angle $\omega$, we have calculated the decay rates for the colorons into various final states:
\begin{align}
\begin{split}\label{BR-eqs}
&\Gamma(C\to jj)=\frac{1}{6\pi}g_S^2 M_C \tan^2\omega \\
&\Gamma(C\to t\bar{t})=\frac{1}{24\pi}g_S^2 M_C \cot^2\omega\sqrt{1-4\frac{m^2_t} {M^2_C}}\left(1+2\frac{m^2_t}{M^2_C}\right) \\
&\Gamma(C\to b\bar{b})=\frac{1}{24\pi}g_S^2 M_C \cot^2\omega \\
&\Gamma(C\to \psi^{1,2}_L\bar{Q}_L)=O(\alpha^2) 
\end{split} 
\end{align}
where $j=u,d,c,s$.  Note that a coloron has no tree-level three-point coupling to gluons. In the following we will neglect $O(\alpha^2)$ terms and we will assume $M_C<2M$; therefore we will be considering only the coloron's decays into ordinary quarks.  Fig. \ref{BR-C} shows the coloron decay branching ratios as a function of $\cot\omega$, for $M_C=1$ TeV.
Fig. \ref{width-C} shows the coloron total decay width as function of $\cot\omega$, for $M_C=1,2,4$ TeV, where $g_S$ has been evaluated at $M_C$.

Each experiment's dataset applies to a certain range of dijet resonance widths, and, therefore, to a certain range of $\cot\omega$ values.
The CMS analysis \cite{cms-pas-exo-12-016} on resonances decaying to dijets applies to narrow resonances, with a half-width smaller than the CMS resolution. Since this resolution is $\sim$ 5 $\%$, the CMS limits will apply to the coloron in the $\cot\omega$ range [0.8, 1.7] for $M_C=2$ TeV and in the range [0.75, 1.8] for $M_C = 4$ TeV.  The ATLAS analysis encompasses broader dijet resonances and therefore probes a wider range of $\cot\omega$, approximately $0.5 \leq \cot\omega \leq 2.5$.
 
We have used MADGRAPH to calculate the cross sections for the process $pp\to C^a\to jj$ in this model at the 8 TeV LHC as function of $M_C$ and for different $\cot\omega$ values.  Moreover, we have multiplied our theoretical cross-sections by the appropriate acceptance values for the coloron signal in the ATLAS and CMS analyses. For the ATLAS analysis, we have calculated the acceptances by following the procedure described in \cite{atlas-1108.6311}, in which we have considered the same kinematic cuts applied in the ATLAS analysis (detailed in \cite{atlas-conf-2012-088}), and we have obtained an acceptance value of 0.44, independent of the coloron mass in the range [1.5 TeV, 4.5 TeV]. For the CMS analysis, we have taken into account the same acceptance value of 0.6, independent of the coloron mass, that was employed in the CMS analysis. 

\begin{figure}
\includegraphics[width=0.45\textwidth]{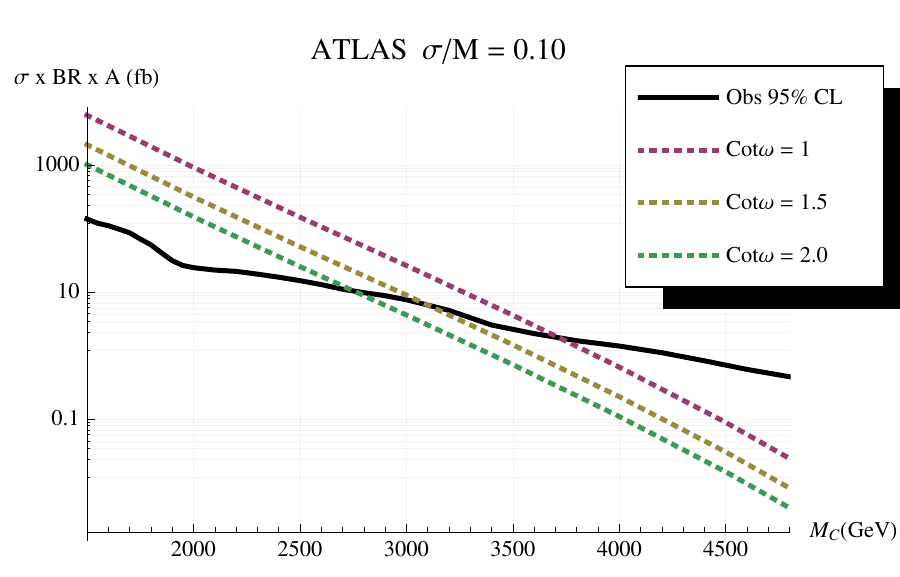}
\includegraphics[width=0.45\textwidth]{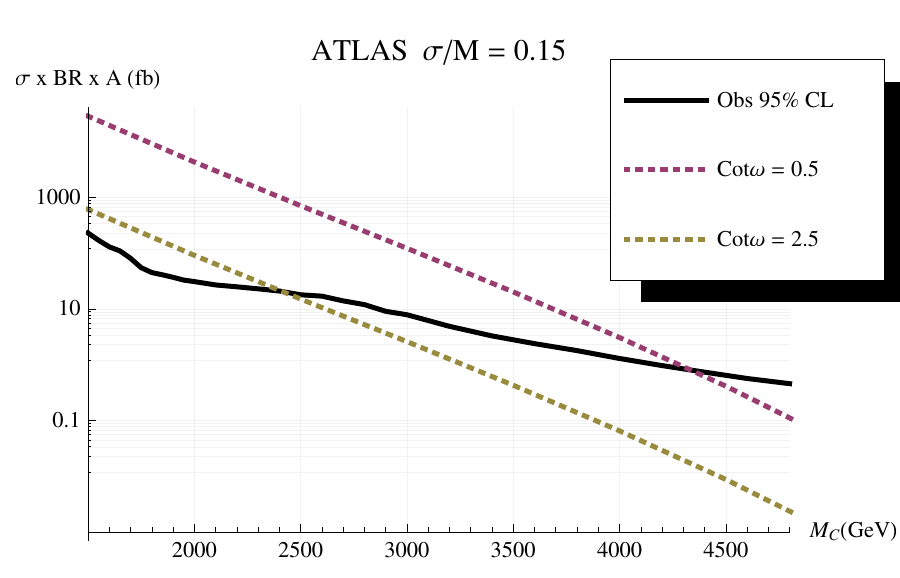}
\caption{Cross-section times branching ratio times acceptance for $pp\to C\to jj$ at the 8 TeV LHC as function of $M_C$ for different values of $\cot\omega$ (dashed colored curves) and the observed $95\%$ C.L. bounds from the ATLAS analysis \cite{atlas-conf-2012-148} on Gaussian resonances (solid black curve). The left-hand plot is for $\sigma/M=0.10$; the right-hand plot is for $\sigma/M=0.15$.}
\label{Atlas-mass-limits}
\end{figure}

\begin{figure}
\includegraphics[width=0.5\textwidth]{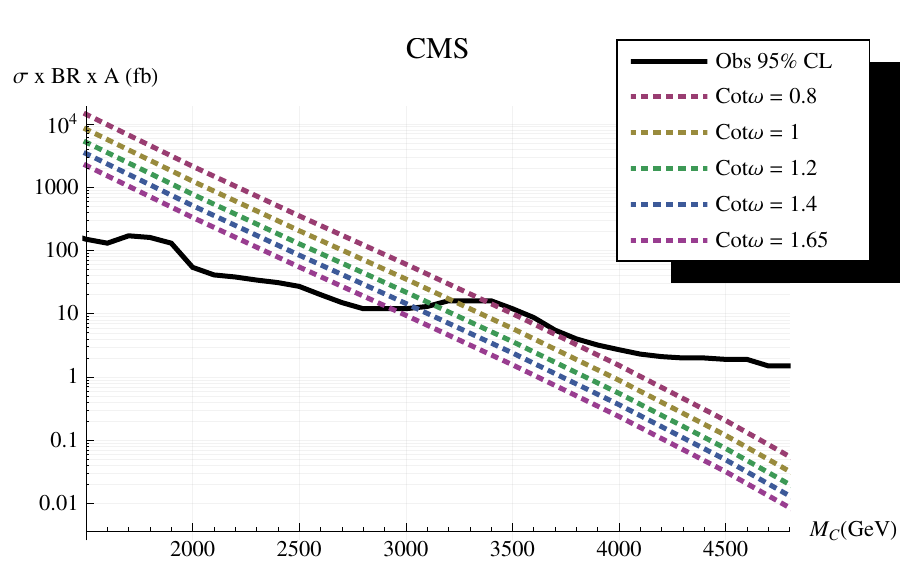}
\caption{Cross-section times branching ratio times acceptance for $pp\to C\to jj$ at the 8 TeV LHC as function of $M_C$ for different values of $\cot\omega$ (dashed colored curves) and the observed $95\%$ C.L. bounds from the CMS analysis \cite{cms-pas-exo-12-016} (solid black curve).}
\label{cms-mass-limits}
\end{figure}

Fig. \ref{Atlas-mass-limits}  compares our theoretical curves (including acceptance factors) to the observed $95\%$ C.L. bounds from the ATLAS analysis \cite{atlas-conf-2012-148} of Gaussian resonances decaying to dijets with $\sigma/M=0.10$ (left) and $\sigma/M=0.15$ (right).   Likewise, Fig. \ref{cms-mass-limits} shows the theory cross-section curves compared with the CMS data from \cite{cms-pas-exo-12-016}. As noted earlier, the ATLAS analysis is sensitive to a broader range of $\cot\omega$ values, but the two data sets give quite consistent results in their region of overlap.  We see that the lower bound on the coloron mass ranges from $M_c \geq 2.4$ TeV for $\cot\omega \approx 2.5$, when the coloron couples mainly to third-generation quarks, all the way to $M_c \geq 4.3$ TeV for $\cos\omega \approx 0.5$, when the coloron couples more strongly to first and second generation quarks.  This bound is overlaid on those from FCNC in Fig. \ref{fig:coupling-plane-sketch}.

\begin{figure}
\includegraphics[width=0.5\textwidth]{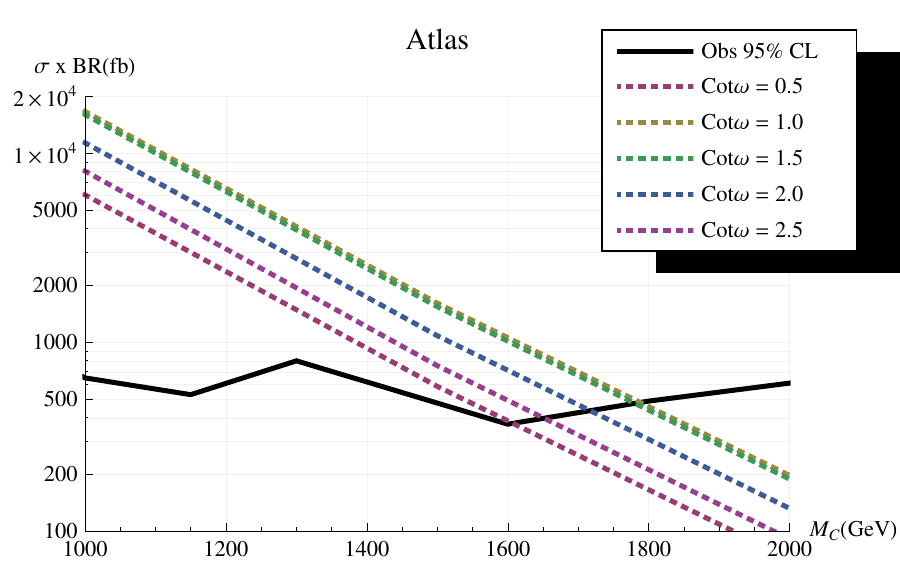}
\caption{Cross-section times branching ratio times acceptance for $pp\to C\to t\bar{t}$ at the 7 TeV LHC as function of $M_C$ for different values of $\cot\omega$ (dashed colored curves) and the observed $95\%$ C.L. bounds from the ATLAS analysis \cite{Atlas-tt} (solid black curve).}
\label{fig-Atlas-tt}
\end{figure}

Finally, we note that the ATLAS \cite{Atlas-tt} and CMS \cite{cms-tt} searches for  resonances decaying to $t\bar{t}$ put weaker bounds on the coloron mass than the searches for resonances in dijets. Fig. \ref{fig-Atlas-tt} compares theoretical curves for the coloron decay into $t\bar{t}$ with the observed $95\%$ C.L. bounds from the ATLAS analysis \cite{Atlas-tt} of the production of a massive KK-gluon (with $\Gamma/M=15\%$) decaying to $t\bar{t}$. 

\section{Vector fermion phenomenology}

LHC data also provides a lower bound on the masses of the heavy quark states that are mostly composed of the vector fermions.  Eq. (\ref{eq:4-3mix}), or equivalently the mixing between vector fermions and $t_R$, $b_R$ SM quarks described in Eqs. (\ref{eq:Tright}) and (\ref{eq:Bright}), implies the following interactions of the $Q=(T,B)$ vector fermions with the electroweak gauge bosons:

\begin{align}
\begin{split}\label{eq:4th-interact}
& -\frac{g}{\sqrt{2}}\lambda^{'}_b \bar{T}_R \gamma^{\mu} W^{+}_{\mu}b_R-\frac{1}{2}\frac{g}{\cos\theta_W}\lambda^{'}_t \bar{T}_R \gamma^{\mu} Z_{\mu}t_R+\frac{\sqrt{2}M}{v}\lambda^{'}_t\bar{T}_L h t_R \\
& -\frac{g}{\sqrt{2}}\lambda^{'}_t \bar{B}_R \gamma^{\mu} W^{-}_{\mu}t_R+\frac{1}{2}\frac{g}{\cos\theta_W}\lambda^{'}_b \bar{B}_R \gamma^{\mu} Z_{\mu}b_R+\frac{\sqrt{2}M}{v}\lambda^{'}_b\bar{B}_L h b_R~.
\end{split}
\end{align}
\noindent
In the limit $\lambda^{'}_b \ll \lambda^{'}_t$, which is favored by constraints on the process $b \to s \gamma$, we see that mass-eigenstate heavy fermions decay into a weak boson ($W$, $Z$, or $h$) plus a right-handed nearly-standard fermion $({t}_R, {b}_R)$ with the following branching ratios:
\begin{align}
\begin{split}\label{eq:4th-BR}
& BR({\mathsf B}\to W {\mathsf t}_R)\simeq 1 \\
& BR({\mathsf T}\to Z {\mathsf t}_R)\simeq BR({\mathsf T}\to h {\mathsf t}_R)\simeq 0.5~.
\end{split}
\end{align}
The last equality arises because the decay ${\mathsf T}\to Z {\mathsf t}_R$ is largely to longitudinally polarized $Z$-bosons, and the equality is a consequence of the Equivalence Theorem.

The heavy $\mathsf B$, $\mathsf T$ can be produced at the LHC in pairs via gluon-gluon fusion \cite{Contino:2008hi, AguilarSaavedra:2009es,Dissertori:2010ug} or singly, through their interactions with $W$, $Z$, or $h$ as in Eq. (\ref{eq:4th-interact}) \cite{AguilarSaavedra:2009es, Mrazek:2009yu, Vignaroli:2012sf}. The ATLAS search in these channels \cite{Atlas-conf-130} for a 4th generation down-type quark, which decays predominantly into $Wt$, puts a limit on the 4th generation quark mass that we can directly apply to the ${\mathsf B}$ vector fermion mass:

\begin{equation}
M_{\mathsf B}\gtrsim 0.67\ \text{TeV}
\label{eq:MBlim}
\end{equation}    
\noindent
Analogously, the $\mathsf T$ vector fermion can be discovered at the LHC through its double production in the final states: $ZZt\bar{t}$, $Zht\bar{t}$, $hht\bar{t}$ or via its single production in the final states: $Zt\bar{t}+jets$, $ht\bar{t}+jets$. The CMS search for a vector-like charge-2/3 quark that decays predominantly into $Zt$ \cite{Chatrchyan:2011ay}, yields a somewhat milder constraint on $M_{\mathsf T}$,  \begin{equation}
M_{\mathsf T}\gtrsim 0.475\ \text{TeV}\,.
\label{eq:MTlim}
\end{equation}  

\section{Conclusions and Outlook}

We have introduced a simple renormalizable model based on an extended color gauge sector that couples differently to the third generation than to the lighter-generation quarks. In addition to the usual SM gauge bosons, there is also a color-octet of top-colorons that single out the third generation quarks.  Mixing between the third-generation of quarks and the first two is naturally small, and occurs only through the (suppressed) mixing of all three generations of ordinary quarks with a set of heavy weak-vector quarks.  Because the third generation and vector quarks transform under one $SU(3)$ group and the light quarks transform under the other, the pattern of quark masses and CKM mixings is reproduced naturally under the conditions summarized in Eq.~(\ref{est-ckm}).

Moreover, this flavorful top-colorn model, which exemplifies next-to-minimal flavor violation, is also consistent with current experimental limits from FCNC, searches for new dijet or top-pair resonances, or searches for new heavy fermions.  Fig. \ref{fig:Excl-lambda-tb} illustrates the range of Yukawa coupling space that is consistent with $b\to s\gamma$, while Fig.~\ref{fig:coupling-plane-sketch} shows how limits from $b\to s\gamma$, neutral meson mixing, and dijet resonance searches restrict the mass and coupling of the top-colorons.  

Not only is this model consistent with current data, but it also offers promising avenues for future exploration at the LHC.  Present limits tell us that the top-coloron mass must be in the TeV range, while the new heavy vector quarks must have masses greater than 670 GeV.  These values leave the new colored states well within the range of the LHC's upcoming high energy run.  We look forward to seeing what the experiments will discover!

\begin{acknowledgments}
R.S.C., E.H.S., and \ N.V. were supported, in part, by the U.S.\ National Science Foundation under Grant No.\ PHY-0854889. R.S.C. and E.H.S. thank the Galileo Galilei Institute for Theoretical Physics for the hospitality and the INFN for partial support during the completion of this work.

\end{acknowledgments}

\appendix

\section{Derivation of bound on $C_7$}
\label{App_C7_bound}
The Standard Model prediction and the experimental measurement \cite{exp-bsgamma} of the $b\to s \gamma$ branching ratio are respectively:
\begin{equation}
 BR_{th}=(315 \pm 23)10^{-6}
 \label{BRth}
\end{equation}
 \begin{equation}
  BR_{ex}=(343 \pm 21 \pm 7)10^{-6}
 \label{BRex}
\end{equation}
The $b\to s \gamma$ decay rate, including both Standard Model (SM) and New Physics (NP) contributions is:
\begin{equation}
\Gamma_{tot} \propto |\mathcal{C}_7(\mu_b)|^2+|\mathcal{C}^{'}_7(\mu_b)|^2\approx 
|\mathcal{C}^{SM}_7(\mu_b)+\mathcal{C}^{NP}_7(\mu_b)|^2+|\mathcal{C}^{'NP}_7(\mu_b)|^2
\label{gamma_tot}
\end{equation}
\noindent
If we consider only the $\mathcal{C}_7$ contribution (since we have found the $C'_7$ piece to be suppressed in our model), we obtain:
\begin{equation}
\frac{\Gamma_{tot}}{\Gamma_{SM}}=1+2\frac{\rm Re(\mathcal{C}^{SM}_7(\mu_b)^{*}\mathcal{C}^{NP}_7(\mu_b))}{|\mathcal{C}^{SM}_7(\mu_b)|^2}+ \mathcal{O}([\mathcal{C}^{NP}_7]^2)\,.
\label{gamma_ratio}
\end{equation}
For $\mu_b =5$ GeV, $\mu_W = M_W $, $\alpha_S \equiv (g_S^2 / 4\pi) = 0.118$, the SM contribution to $C_7$ at the scale $\mu_b$ reads \cite{Buras}:

\begin{equation}
C^{SM}_7(\mu_b)=0.695 C^{SM}_7(\mu_W)+0.086 C^{SM}_8(\mu_W)-0.158 C^{SM}_2(\mu_W)=-0.300 \ .
\label{C7SM}
\end{equation}
The scaling factor of the New Physics contribution to $C_7$ from the scale $\mu_W$ to the scale $\mu_b$ is:

\begin{equation}
 \mathcal{C}^{NP}_7(\mu_b) = \left(\frac{\alpha_S (\mu_W)}{\alpha_S (\mu_b)} \right)^{\frac{16}{23}}\mathcal{C}^{NP}_7(\mu_w)= 0.695\ \mathcal{C}^{NP}_7(\mu_w) \ ;
\label{C7NPscale}
\end{equation}
\noindent
that from the scale $m_* = 1$ TeV  to the scale $\mu_W$ is:
\begin{equation}
 \mathcal{C}^{NP}_7(\mu_W)=\left(\frac{\alpha_S (m_*)}{\alpha_S (m_t)} \right)^{\frac{16}{21}} \left(\frac{\alpha_S (m_t)}{\alpha_S (\mu_W)} \right)^{\frac{16}{23}}\simeq 0.79\ \mathcal{C}^{NP}_7(m_*) \
 \label{scale_mStar}
\end{equation}
\noindent
and we obtain at $95 \%$ C.L.:
\[
 -0.093 < \rm Re  [ \mathcal{C}^{NP}_7(m_*)] < 0.023
\]

 \end{document}